\documentclass[twocolumn,showpacs,preprintnumbers,amsmath,amssymb]{revtex4}

\usepackage{graphicx}
\usepackage{dcolumn}
\usepackage{bm}
\usepackage{color}
\usepackage{comment}
\usepackage{here}

\begin{document}

\preprint{APS/123-QED}

\title{Emergence of the minority hole with high mobility on the electrical transport in the Fe-pnictides Ba(Fe$_{1-x}$Mn$_x$As)$_2$}

\author{T. Urata$^1$}

\author{Y. Tanabe$^1$}
\thanks{Corresponding author: ytanabe@m.tohoku.ac.jp}

\author{K. K. Huynh$^2$}

\author{S. Heguri$^2$}

\author{H. Oguro$^3$}

\author{K. Watanabe$^3$}

\author{K. Tanigaki$^{1, 2}$}
\thanks{Corresponding author: tanigaki@sspns.phys.tohoku.ac.jp}

\affiliation{$^1$Department of Physics, Graduate School of Science, Tohoku University, Aoba, Aramaki, Aoba-ku, Sendai, 980-8578, Japan}

\affiliation{$^2$WPI-Advanced Institutes of Materials Research, Tohoku University, Aoba, Aramaki, Aoba-ku, Sendai, 980-8578, Japan}

\affiliation{$^3$High Field Laboratory for Superconducting Materials, Institute for Materials Research, Tohoku University, Sendai 980-8577, Japan}

\date{\today}

\begin{abstract}
In Fe pnictide (Pn) superconducting materials, neither Mn- nor Cr- doping to the Fe site induces superconductivity, even though hole carriers are generated.
This is in strong contrast with the superconductivity appearing when holes are introduced by alkali metal substitution on the insulating blocking layers.
We investigate in detail the effects of Mn doping on magneto-transport properties in Ba(Fe$_{1-x}$Mn$_x$As)$_2$ for elucidating the intrinsic reason. 
The negative Hall coefficient for $x$ = 0 estimated in the low magnetic field ($B$) regime gradually increases as $x$ increases, and its sign changes to a positive one at $x$ = 0.020.
Hall resistivities as well as simultaneous interpretation using the magnetoconductivity tensor including both longitudinal and transverse transport components clarify that minority holes with high mobility are generated by the Mn doping via spin density wave (SDW) transition at low temperatures, while original majority electrons and holes residing in the parabolic-like Fermi surfaces (FSs) of the semimetallic Ba(FeAs)$_2$ are negligibly affected.
Present results indicate that the mechanism of hole doping in Ba(Fe$_{1-x}$Mn$_x$As)$_2$ is greatly different from that of the other superconducting FePns family.
\end{abstract}

\maketitle

\section{INTRODUCTION}
In high temperature superconducting (HTS) materials, elemental substitutions induce many intriguing changes in physical properties.
In HTS-cuprates, an elemental substitution to an insulating blocking layer provides additional mobile carriers in the Mott insulating CuO$_2$ plane, resulting in the well-known high temperature superconductivity \cite{LBCO}.
To be contrast with the substitution effect on the insulating blocking layer, doping of both nonmagnetic and magnetic impurities to the conducting CuO$_2$ layer strongly suppresses the superconductivity via the unconventional superconducting mechanism \cite{Tc, PB}, resulting in magnetism \cite{stripe} and carrier localization \cite{Htrap}. 
On the other hand, in HTS Fe-pnictides (FePns), although elemental substitution to the insulating blocking layer induces the high temperature superconductivity\cite{1111F, 1111H, BaK122, CaRE122}, the substitution effects on the Fe-Pn conducting layers are quite complex due to the Fe-3$d$ multiband nature.
The transition metal (TM) doping on the Fe site by Co, Ni, Rh, Pd, and Pt, located at the right position of Fe in the periodic table, suppresses the colinear-type spin density wave (SDW) and induces superconductivity \cite{BaCo122,BaNi122,BaRh_Pd,BaPt122}.
A rigid band like upper shift of the chemical potential observed by angle resolved photo emission spectroscopy (ARPES) \cite{Sekiba,Brouet,Neupane} gives an evidence of electron doping to the FeAs layer.
Further extended studies on the phase diagram proposed that the superconducting transition temperature ($T_{\rm c}$) well scales both with the structural and the magnetic transition temperatures ($T_{\rm S}$ and $T_{\rm N}$, respectively) in the underdoped region and with the number of extra electrons added by the TM doping in the overdoped region of the superconducting dome \cite{Canfield_rev}.
On the other hand, recent ARPES studies on Ba(Fe$_{1-x}$M$_x$As)$_2$ (M = Ni, Cu) demonstrated that the number of electrons does not monotonically increase with extra $d$-electrons introduced per Fe/TM site \cite{Ideta}.
In the case of TMs located at the left of Fe, doping of either Mn or Cr does not induce any superconductivity and instead stabilizes the checkerboard-like magnetic order \cite{Tucker}.
A clear sign change of the $R_{\rm H}$ by Mn and Cr substitution indicated a fact that holes are doped \cite{Sefat,Urata}, while nuclear magnetic resonance (NMR), neutron scattering, photoemission spectroscopy (PES) and x-ray absorption spectroscopy (XAS) demonstrated the localization of Mn 3$d$ states \cite{Tucker,Texier,Suzuki}.
The intrinsic effects of TM doping on the Fe-Pn conducting layer are still controversial.
Since intrinsic hole doping to the Fe-Pn layers should produce superconductivity, the sign change in $R_{\rm H}$ for non-superconducting Ba(Fe$_{1-x}$M$_x$As)$_2$ (M = Mn, Cr) is important for understanding the role of the TM doping on the superconducting phase diagram of HTS FePns.

In this paper, we show the emergence of the minority holes with high mobility based on the detailed analyses of electrical transport in Ba(Fe$_{1-x}$Mn$_x$As)$_2$.
Importantly, minority holes with high mobility are shown to newly emerge in addition to the majority carriers with relatively low mobility.
Simultaneous analyses of both longitudinal and transverse magnetoconductivities ($\sigma_{xx}$ and $\sigma_{xy}$, respectively) consistently show the emergence of minority holes with high mobility, in addition to the majority electrons and holes with low mobility residing in the parabolic bands that were reported in the parent compound as well.
Since the Mn 3$d$ states are nearly localized in Ba(Fe$_{1-x}$Mn$_x$As)$_2$ \cite{Tucker,Texier,Suzuki}, it is likely that the minority holes are generated by the Mn doping with keeping the original semimetallic band structure of Ba(FeAs)$_2$ \cite{Urata}.
Present results demonstrate the considerable large contribution of the minority carriers with high mobility to the magneto-transport properties for having the true understanding of the TM doping in the electronic phase diagram of HTS FePns.

\section{EXPERIMENTAL DETAILS}
Single crystals of Ba(Fe$_{1-x}$Mn$_x$As)$_2$ with $x =0, 0.007, 0.016 ,0.020, 0.035$ were grown by a flux method using FeAs \cite{Canfield}.
Synchrotron X-ray diffraction measurements (SPring8 BL02B2) and Rietveld refinements (by GSAS and its graphical user interface EXPGUI \cite{GSASandEXPGUI}) were performed to check the quality of the single crystals.
Temperature dependencies of electrical resistivity($\rho$) were also measured to deduce the $T_{\rm N}$s (Fig. 1 (a)).
In figure 1 (b), the first derivative of a normalized resistivity curve (d($\rho /\rho_{\rm 300K}$)/d$T$) showed a clear inflection point at $T_{\rm N}$, being consistent with the previous reports \cite{Canfield, Urata}.
The Mn concentration $x$ was determined using the $x$ dependence of $T_{\rm N}$ for Ba(Fe$_{1-x}$Mn$_x$As)$_2$ previously reported \cite{Canfield}.
Both in-plane transverse magnetoresistance ($R_{\rm M}$) and Hall resistance were measured using a four-probe method for $|B| \leq 18$ T at various temperatures between 2 and 180 K.
It should be noted that the magnetoresistance (the Hall resistance) is averaged (subtracted) between positive and negative field to cancel antisymmetric (symmetric) factor.

\begin{figure}[t]
\includegraphics[width=0.8\linewidth]{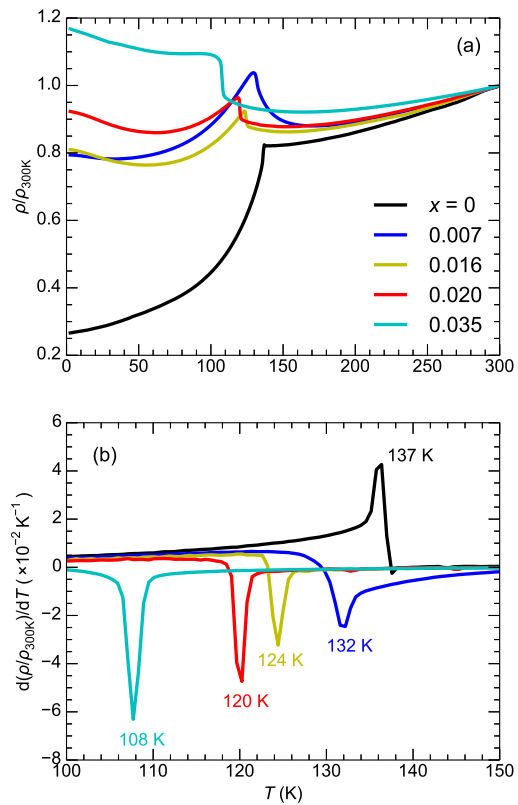}
\caption{Temperature dependence of (a) normalized resistivity ($\rho /\rho_{\rm 300K}$) and (b) first derivative of $\rho /\rho_{\rm 300K}$ for Ba(Fe$_{1-x}$Mn$_x$As)$_2$. Magnetic transition temperatures ($T_{\rm N}$s) are derived from infliction points of resistivity curves.}
\end{figure}


\section{RESULTS}
\subsection{Evolution of the Hall resistivity with the tiny Mn doping in Ba(Fe$_{1-x}$Mn$_x$As)$_2$}

\begin{figure}[t]
\includegraphics[width=0.8\linewidth]{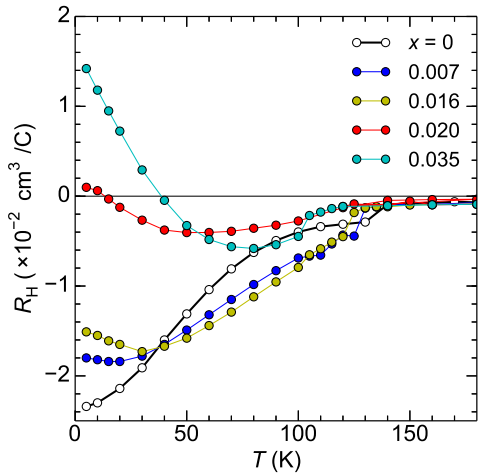}
\caption{Temperature dependence of Hall coefficients ($R_{\rm H}$s) for Ba(Fe$_{1-x}$Mn$_x$As)$_2$ at low magnetic field ($B$) limits. $R_{\rm H}$s are derived by a linear fitting of a $\rho_{yx}$ curvature where it develops linearly at low $B$ regime as shown in Fig. 3.}
\end{figure}

\begin{figure}[t]
\includegraphics[width=0.8\linewidth]{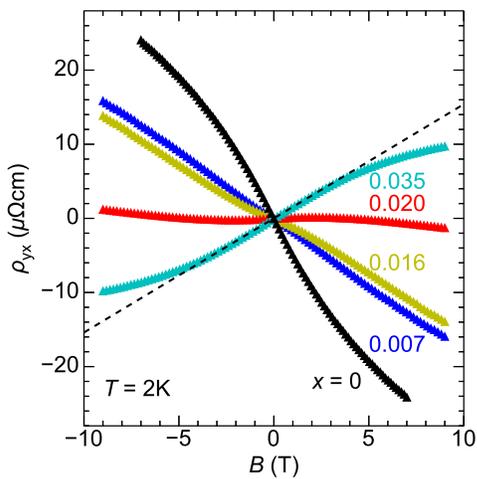}
\caption{(a) Magnetic fields ($B$s) dependence of Hall resistivity ($\rho_{yx}$) for Ba(Fe$_{1-x}$Mn$_x$As)$_2$ with various concentration $x \leq 0.035$ at $T = 2$ K. A broken line indicates an example of a linear fitting deriving the Hall coefficients ($R_{\rm H}$s) at low $B$ regime.}
\end{figure}

\begin{figure}[t]
\includegraphics[width=1.0\linewidth]{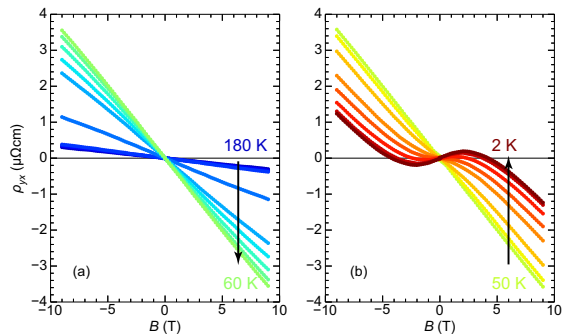}
\caption{(a) Magnetic fields ($B$s) dependence of Hall resistivity ($\rho_{yx}$) for Ba(Fe$_{1-x}$Mn$_x$As)$_2$ with $x$ = 0.020 at various temperature ($T$) ((a)$T$ = 60,70,80,90,100,120,140,160,180 K and (b)$T$ = 2,5,10,15,20,30,40,50 K). Arrows indicate the direction of $T$ evolution from high- to low-$T$.}
\end{figure}

Fig. 2 shows the temperature dependence of $R_{\rm H}$ for Ba(Fe$_{1-x}$Mn$_x$As)$_2$ with $0 \leq x\leq 0.035$.
The $R_{\rm H}$s were derived by linear fitting of $\rho_{yx}$ curves at low $B$ where they develop linearly as shown in Fig. 3.
The $R_{\rm H}$s are negative and almost constant at temperatures above $T_{\rm N}$.
Below each $T_{\rm N}$, the $R_{\rm H}$s show large variations against temperature.
For $x$ = 0.007, the $R_{\rm H}$ decreases with a decrease in temperature.
Around 20 K, it shows a local minimum and increases with a decrease in temperature.
This increase in $R_{\rm H}$ at low temperatures gradually becomes more remarkable with increasing $x$ and finally gives a positive $R_{\rm H}$ at $x \geq 0.020$.
For elucidating the reason of this positive $R_{\rm H}$ in details, the $B$ dependence of Hall resistivity ($\rho_{yx}$) curves is useful.
Fig. 3 shows $B$ dependencies of the $\rho_{yx}$ for $x = 0, 0.007, 0.016, 0.020, 0.035$ at 2 K.
With increasing $x$, the overall slope of $\rho_{yx}$ gradually changes sign from negative to positive.
Intriguingly for $x$ = 0.020, the sign of $\rho_{yx}$ gradually changes from positive to negative with an increase in $B$.
In order to clarify this complex phenomenon observed in $\rho_{yx}$ of the $x$ = 0.020 sample, the $B$ dependence of $\rho_{yx}$ at various temperatures are displayed in Fig.4.
The linear shape in $\rho_{yx}$s at high temperatures gradually changes to convex one at low temperatures below 50 K under low $B$.
On the other hand, under high-$B$ similar curvatures with negative values of $\rho_{yx}$ are observed in the entire temperature range between 2 and 180 K.
This experimental observation is intriguing and indicative to the emergence of minority holes with high mobility under the compensating nature of majority FSs, as will be discussed later.

In the semiclassical transport theory, the $R_{\rm H}$ under low $B$ is described as \cite{AM},
\begin{eqnarray}
R_{\rm H} &=& \frac{\sum^N_{i=1}a_in_i\mu_i^2}{e(\sum^N_{i=1}n_i\mu_i)^2},
\end{eqnarray}
where $\mu_i$ and $n_i$ are the carrier mobility and the carrier density of the $i$-th band, respectively.
$a_i$ takes the value of +1 for hole and -1 for electron.
In the semimetallic electronic structures such as FePns, the electron and hole compensation for low $\mu$ bands significantly suppresses the absolute value of $R_{\rm H}$.
Moreover, carriers with high $\mu$ give a significant contribution to $R_{\rm H}$.
Therefore, $R_{\rm H}$ at low $B$ is significantly influenced by the minority carriers with high mobility, while at high $B$ it is dominated by the majority carriers with low mobility.
This point described here can provide preliminary qualitative interpretation for $\rho_{yx}$ in the framework of semiclassical transport theory.
The positive $\rho_{yx}$ slope under low $B$ clearly demonstrates the existence of minority holes with high mobility for $x$ = 0.020 instead of the minority electrons with high mobility in Ba(FeAs)$_2$ \cite{KhuongNJP}, being markedly different from what have been observed for Co and Ni so far.
The negative $\rho_{yx}$ slope observed under high $B$ indicates that FSs including majority electron carriers with relatively lower mobility also residing behind the present experimental data.
Consequently, a reasonable interpretation to explain the two extreme limits of $B$ in the $\rho_{yx}$ behavior for $x$ = 0.020 is given as follows:
In electric transport, the minority holes with high mobility are gradually pronounced with an increase in $x$ in the presence of the majority FSs as a background, and this scenario is also consistent with the gradual increase in $\rho_{yx}$ with an increase in $x$ at low temperatures.

It can be found from the detailed Hall resistivity of Ba(Fe$_{1-x}$Mn$_x$As)$_2$ in the present study that Mn doping generates minority holes with high mobility in addition to the majority FSs, giving a large influence on R$_{H}$.
Since previous reports indicated localization of the Mn 3$d$ states \cite{Texier,Tucker,Suzuki}, Mn doping is considered to provide merely a small influence on both the majority electron and hole FSs composed of the original $d$-orbitals of Ba(FeAs)$_2$.
When all discussions described so far are taken into consideration, the following scenario for the electronic structure of Ba(Fe$_{1-x}$Mn$_x$As)$_2$ can be given:
The Mn doping generates the minority holes with high mobility instead of the minority electrons with high mobility in Ba(FeAs)$_2$ \cite{KhuongNJP}.
The carrier number of the majority electron and the hole FSs hold almost constant when Mn is doped into the Fe site, but these contributions to the electric transport become apparent due to the significant contribution in electric transport by the minority holes with high mobility newly appearing as temperature is lowered.

\subsection{Simultaneous multicarrier analysis for longitudinal and transverse magneto-conductivity in Ba(Fe$_{1-x}$Mn$_x$As)$_2$}

\begin{figure*}[htbp]
\includegraphics[width=0.7\linewidth]{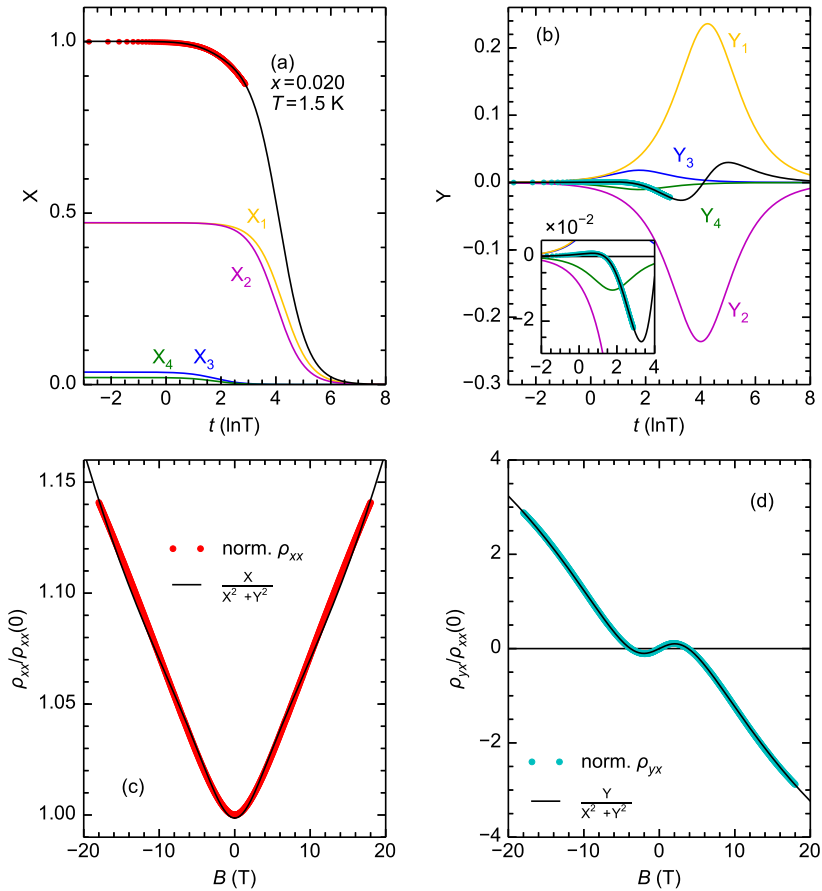}
\caption{Simultaneous multicarrier analysis of longitudinal and transverse magneto-conductivity in Ba(Fe$_{1-x}$Mn$_x$As)$_2$ with $x$ = 0.020 at $T = 1.5$ K. 
(a), (b) Normalized longitudinal and transverse conductivity ($X$ and $Y$, respectively). Black lines are results of fitting using the four carrier model. 
(c), (d) Normalized longitudinal and transverse resistivity ($\rho_{xx}/\rho_{xx}(0)$ and $\rho_{yx}/\rho_{xx}(0)$, respectively). 
Black lines are derived from fitting functions by applying the general relationship between resistivity and conductivity tensors (i.e. $\rho = \sigma^{-1}$).}
\end{figure*}

In order to confirm the electronic structure deduced from the Hall resistivity studies, here we carried out the analyses of the magnetoconductivity.
Fig. 5(a) and (b) show the normalized longitudinal and transverse conductivity ($X$ and $Y$, respectively),
where $X$ and $Y$ are defined as $X = \sigma_{xx}(B)/\sigma_{xx}(0)$ and $Y = \sigma_{xy}(B)/\sigma_{xx}(0)$.
In a logarithmic scale, the fitting functions $X$ and $Y$ can be written as \cite{JSKim},
\begin{eqnarray}
    X_j(t) &=&  \frac{|f_j|}{1+e^{2(t-m_j)}} \label{4}\\
    Y_j(t) &=&  \frac{f_j}{2{\rm cosh}(t-m_j)} \label{5},
\end{eqnarray}
where $t \equiv {\rm ln}B$, $m_j \equiv -{\rm ln}\mu_j$ and $f_j \equiv \pm{\rm e}N_j\mu_j/\sigma_{xx}(0)$ (positive for hole and negative for electron).
Here, e, $N_j$, and $\mu_j$ denote the electron charge, the carrier density, and the carrier mobility of the $j$-th band, respectively.
Since the same parameters are commonly used in the two formulae, we must fit both conductivities simultaneously to obtain the solution from one experimental dataset.
The first equation is a step-like function and the second is a peak-like function of $t$.
The $B$ positions of both the step in $X_j$($B$) and the peak in $Y_j$($B$) represent the carrier mobility (i.e. if $t$=$m_j$, then $B$=$1/\mu$).
The step and the peak heights reflect the carrier number.
These representations of magnetoconductivities are useful for understanding the electronic states.
The whole fitting functions are represented as $\sum X_j$ and $\sum Y_j$.
The summation should be taken up to the correct number of carrier types.
The band calculations proposed the two electron- and two hole- FSs in the colinear-type SDW Ba(FeAs)$_2$ \cite{Yin}.
Moreover, the Mn 3$d$ states are localized \cite{Texier,Tucker,Suzuki} and the colinear-type SDW phase is conserved below $x$ = 0.10 \cite{MGKim}.
Accordingly, it is reasonable to employ a four-carrier model for the present analysis of the magneto-conductivity \cite{Yin,KhuongNJP}.

The predominant contribution in semimetallic electronic structure of the present compounds results from the large electron and hole FSs (conventional parabolic bands).
In the analyses, however, two additional small FSs (Dirac cones) to give the minority carriers should be taken into account.
The latter two types minority carriers give nonnegligible contributions to the electric transport due to their much higher mobility.
As shown in Fig. 5 (a) and (b), both $X$ and $Y$ curves for $x$ = 0.020 were successfully reproduced using four types of carrier bands ($\sum X_j$ and $\sum Y_j$ ($j$ = 1, 2, 3, 4)) in the range of $B$=0 to 18 T.
It should be noted that the simultaneous multicarrier fitting using less than four carrier types with any parameters did not reproduce the experimental magnetoconductivities observed in the present studies.
The normalized longitudinal and transverse magnetoresistivities ($\rho_{xx}/\rho_{xx}(0)$ and $\rho_{yx}/\rho_{xx}(0)$, respectively) are shown in fig. 5 (c) and (d) to visually demonstrate the consistency between the fitting curves and the experimental data.
Here fitting curves are obtained from $X$ and $Y$ in the four carrier fitting using the general relationship between resistivity and conductivity tensors (i.e. $\rho = \sigma^{-1}$).
Both $\rho_{xx}/\rho_{xx}(0)$ and $\rho_{yx}/\rho_{xx}(0)$ curves are almost reproduced by the fitting curves, although small errors are observed between the $\rho_{xx}/\rho_{xx}(0)$ and the fitting curve.
A linear component in the magnetoresistance at high $B$ originating from the quantum limit of the Dirac cone \cite{KhuongPRL, Urata} hinders us to achieve a better fitting, which may be one of reasons for the small errors between the $\rho_{xx}/\rho_{xx}(0)$ data and the fitting curves.

\begin{table*}[htbp]
\caption{Results of the simultaneous analysis of both logitudinal and transverse magnetoconductivity for $x$ = 0.020 at $T = 1.5$ K using the four carrier model.
 $j$ indicates the band index.}
\scalebox{0.9}{
\begin{tabular}{l c c c c} \hline
$j$ & 1 & 2 & 3 & 4 \\ \hline \hline
Carrier type & Electron& Hole& Electron & Hole\\ \hline
Mobility (cm$^2$/V$\cdot$s)& 1710$\pm$90 & 1700$\pm$50 & 180$\pm$10 &140$\pm$10\\ \hline
Density (cm$^{-3}$) & 5.39 $\pm$0.60 $\times$10$^{17}$ & 9.48$\pm$0.52 $\times$10$^{17}$ & 1.15$\pm$0.10 $\times$10$^{20}$ & 1.47$\pm$0.14 $\times$10$^{20}$ \\ \hline
\end{tabular}}
\end{table*}

The fitting parameters are displayed in Table I.
The minority electrons and holes ($\sim$ 10$^{17}$ cm$^{-3}$) with high mobility ($\sim$ 1700 cm$^2$(VS)$^{-1}$) as well as the majority electrons and holes ($\sim$ 10$^{20}$ cm$^{-3}$) with relatively low mobility (180 and 140 cm$^2$(Vs)$^{-1}$, respectively) were successfully estimated.
It should be emphasized that the minority electrons and holes with high mobility ($\sim$ 10$^{17}$ cm$^{-3}$ and $\sim$ 1700 cm$^2$(Vs)$^{-1}$) were necessary in order to reproduce $X$ and $Y$ curves in the low $B$ regime.
The majority electrons and holes ($\sim$ 10$^{20}$ cm$^{-3}$) with relatively low mobility (180 and 140 cm$^2$(Vs)$^{-1}$, respectively) provide a situation that an almost compensated electronic structure is realized similarly to the original band structure of Ba(FeAs)$_2$.
The fact that the absolute value of $Y$ is significantly smaller than that of $X$ also supports the above results.
Unfortunately the detailed $X$ and $Y$ behaviors originating from the majority electrons and holes with low mobility in FSs are not apparently clear in the present analyses because of the still insufficient magnetic field strength such as the $X$ ranges of $\sim$ 0.88 at $B$ = 18 T.
We would need extremely high magnetic field in order to clarify the contribution from the majority lower mobility carriers.
Considering the accessible range of magnetic field in our present experiments, although certain errors cannot be ruled out for estimating parameters listed in the Table I, the conclusion that the minority holes with high mobility is necessary to reproduce the $X$ and $Y$ experimental data is not altered.

\section{Discussion}

The simultaneous fitting of the magnetoconductivity data in the framework of semiclassical description showed that the Mn doping generates minority holes with high carrier mobility in Ba(Fe$_{1-x}$Mn$_x$As)$_2$ via SDW transition.
In the SDW state of Ba(FeAs)$_2$, tiny electron-like Dirac pockets were theoretically predicted and also experimentally confirmed by various experiments \cite{Yin, Richard, ImaiJPSJ, KhuongPRL, KhuongNJP}.
Such Dirac cone states are robust for both nonmagnetic and magnetic impurity \cite{Ran, Morinari, Tanabe1, Tanabe2, Urata}, being different from those existing in topological insulators, and consequently the colinear type SDW phase also survives up to $x$ = 0.10.
Therefore, one might expect that hole doping in the electron-like Dirac cone in the parent compound can be realized.
In this case, the minority holes with high mobility may appear as a consequence of generation of hole-like Dirac cones in the SDW Ba(Fe$_{1-x}$Mn$_x$As)$_2$ states by preserving the original majority electron and hole FSs.
Since the carrier mobility of the majority electrons and holes in the conventional parabolic bands are suppressed by the backward scatterings of Mn, the minority Dirac pockets protected by the $\pi$-Berry phase will give rise to significant contributions in the electrical transport, being consistent with the present observations.
Another scenario is that the enhancement of the FS anisotropy.
In Ba(FeAs)$_2$, the electron FS including both convex and concave parts was theoretically and experimentally pointed out \cite{Yin, Shimojima}.
In this case, the cyclotron motion of the electrons at the concave part can induce a hole-like contribution \cite{Ong} and it was actually demonstrated by the mobility spectrum analyses \cite{KhuongNJP}.
The recent report on in-plane electrical resistivity of the detwined Ba(Fe$_{1-x}$Mn$_x$As)$_2$ revealed that the Mn doping increases the anisotropy of the in-plane resistivity \cite{TajimaAPS}.
Therefore, it is also possible that the Mn doping enhances the FS anisotropy (presumably for the tiny electron-like Dirac pocket) in Ba(Fe$_{1-x}$Mn$_x$As)$_2$, resulting in the emergence of the minority holes with high mobility.

\section{Conclusion}
We showed the emergence of the minority holes with the high mobility on the magnetotransport of Ba(Fe$_{1-x}$Mn$_x$As)$_2$.
The negative Hall coefficients estimated in the low magnetic field ($B$) regime gradually increased with an increase in $x$ and changed to a positive one at $x$ = 0.020.
Detailed studies on the Hall resistivities as well as simultaneous analyses of both longitudinal and transverse magnetoconductivities demonstrated that the minority holes with high mobility were generated by the Mn doping, while the majority electron and hole Fermi surfaces (FSs) in the semimetallic electronic structure of parent SDW Ba(FeAs)$_2$ were almost preserved.
The present results elucidated that 
(i) the generated holes as the Dirac pockets in Ba(Fe$_{1-x}$Mn$_x$As)$_2$ are minority from the viewpoint of the total carrier number, 
(ii) the majority FSs in parent Ba(FeAs)$_2$ may not change, and the effect of Mn doping can be distinguished from the pure hole doping to the Fe-Pnictide layer. 
A question is still left about the origin of these minority holes with high mobility.
One possible origin is that Mn induces a small amount of holes to the tiny electron like Dirac pocket of parent Ba(FeAs)$_2$ \cite{Yin, Richard, ImaiJPSJ, KhuongPRL, KhuongNJP}.
Another possibility is that Mn induces the enhancement of the anisotropy of the FSs.
The recent in-plane electrical transport reveals that the in-plane resistivity anisotropy is enhanced by the Mn doping.
A scenario that additional hypothetical carriers can be generated depending on the shape along the segments of the FS \cite{KhuongNJP, Ong} is also applicable.

\section{Acknowledgements}
The authors are grateful to Y. Kuramoto for his useful comments.
The research was partially supported by Scientific Research on Priority Areas of New Materials Science using Regulated Nano Spaces, the Ministry of Education, Science, Sports and Culture, Grant in Aid for Science, and Technology of Japan and Grant-in-Aid for Young Scientists (B) (23740251).
The work was partly supported by the approval of the Japan Synchrotron Radiation Research Institute (JASRI).
One of the authors (T.U.) was supported by the Japan Society for the Promotion of Science.
\section{Appendix}
As noted in the section I\hspace{-.1em}I, the data used in figure 3, 4, and 5 are shown after the manipulation to cancel unnecessary contributions from antisymmetric (symmetric) contribution to the $\rho_{xx}$ ($\rho_{yx}$) which may mostly arise from the misarrangement of the electrical contacts.
The process of the manipulation are written as,
\begin{eqnarray}
    \rho_{xx}&=&\frac{\rho^{\rm raw}_{xx}(|B|)+\rho^{\rm raw}_{xx}(-|B|)}{2}\\
    \rho_{yx}&=&\frac{\rho^{\rm raw}_{yx}(|B|)-\rho^{\rm raw}_{yx}(-|B|)}{2},
\end{eqnarray}

where $\rho_{xx}$ and $\rho_{yx}$ are data after the manipulation (Fig. 2 - 5) and $\rho^{\rm raw}_{xx}$ and $\rho^{\rm raw}_{yx}$ are raw data.
The $\rho^{\rm raw}_{xx}$ and $\rho^{\rm raw}_{yx}$ are shown in figure 6, 7, and 8 corresponding to the figure 3, 4, and 5 before the data manipulation, respectively.

\begin{figure}[h]
\includegraphics[width=0.9\linewidth]{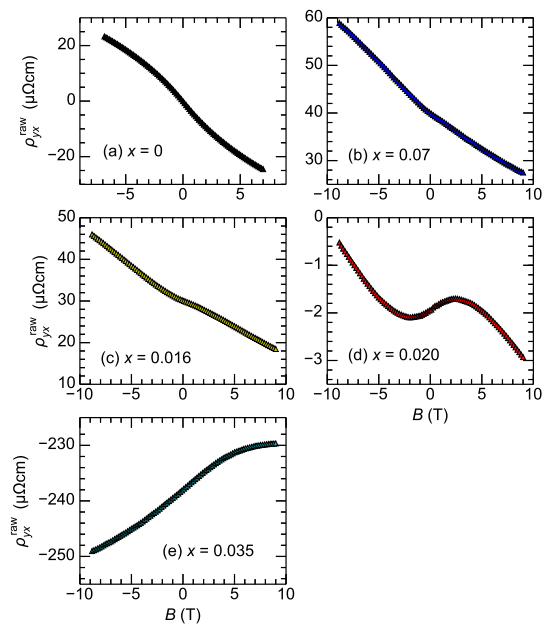}
\caption{The raw data of Hall resistivity ($\rho_{yx}$) magnetic field ($B$) dependence are shown in several concentrations at $T = 2$ K. (a, b, c, d, e)$x = 0, 0.007, 0.016, 0.020, 0.035$, respectively.}
\end{figure}

\begin{figure}[h]
\includegraphics[width=0.9\linewidth]{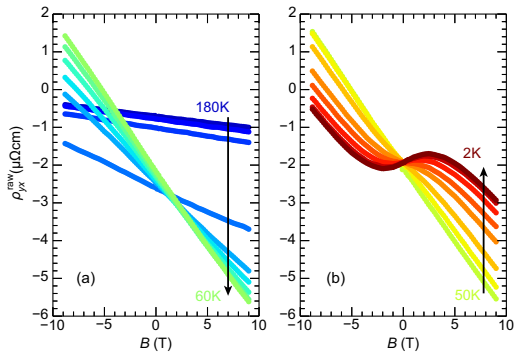}
\caption{The raw data of Hall resistivity ($\rho_{yx}$) magnetic field ($B$) dependence of the $x = 0.020$ sample at various temperatures are shown. ((a)$T$ = 60,70,80,90,100,120,140,160,180 K and (b)$T$ = 2,5,10,15,20,30,40,50 K). Arrows indicate the direction of $T$ evolution from high- to low-$T$.}
\end{figure}

\begin{figure}[h]
\includegraphics[width=0.9\linewidth]{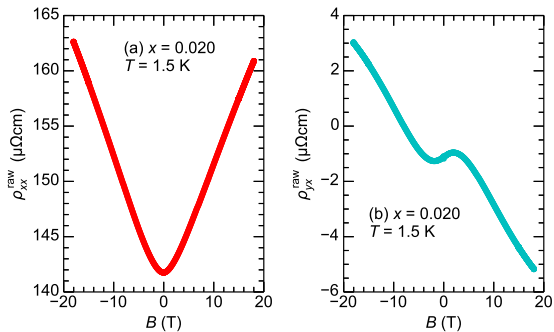}
\caption{The raw data of (a) transverse magnetoresistivity ($\rho_{xx}$) and (b) Hall resistivity ($\rho_{yx}$) magnetic field ($B$) dependence are shown for the $x = 0.020$ sample at $T = 1.5$ K.}
 \end{figure}


\end{document}